\documentclass[aps,prl,superscriptaddress,twocolumn]{revtex4}
\usepackage{graphicx}
\usepackage{amsmath}
\usepackage{bm}
\usepackage{color}

\def\bea{\begin{eqnarray}}
\def\eea{\end{eqnarray}}
\def\be{\begin{equation}}
\def\ee{\end{equation}}

\begin{document}

\title{Precise study of asymptotic physics with subradiant ultracold molecules}

\author{B. H. McGuyer}
\affiliation{Department of Physics, Columbia University, 538 West 120th Street, New York, NY 10027-5255, USA}
\author{M. McDonald}
\affiliation{Department of Physics, Columbia University, 538 West 120th Street, New York, NY 10027-5255, USA}
\author{G. Z. Iwata}
\affiliation{Department of Physics, Columbia University, 538 West 120th Street, New York, NY 10027-5255, USA}
\author{M. G. Tarallo}
\affiliation{Department of Physics, Columbia University, 538 West 120th Street, New York, NY 10027-5255, USA}
\author{W. Skomorowski}
\altaffiliation[Present address:  ]{Institute of Physics, University of Kassel, Heinrich-Plett-Strasse 40, 34132 Kassel, Germany}
\affiliation{Quantum Chemistry Laboratory, Department of Chemistry, University of Warsaw, Pasteura 1, 02-093 Warsaw, Poland}
\author{R. Moszynski}
\affiliation{Quantum Chemistry Laboratory, Department of Chemistry, University of Warsaw, Pasteura 1, 02-093 Warsaw, Poland}
\author{T. Zelevinsky}
\email{tz@phys.columbia.edu}
\affiliation{Department of Physics, Columbia University, 538 West 120th Street, New York, NY 10027-5255, USA}

\begin{abstract}     
Weakly bound molecules have physical properties without atomic analogues, even as the bond
length approaches dissociation.  In particular, the internal symmetries of homonuclear diatomic
molecules result in formation of two-body superradiant and subradiant excited states.  While
superradiance \cite{DickePR54_CollectiveRadiation,EberlyAJP72_Superradiance,HarocheGrossPR82_Superradiance} has been demonstrated in a variety of systems, subradiance \cite{DeVoePRL96_TwoIonsSubradSuperrad,OdomZhouNNano11_SubradiantPlasmons,TakahashiTakasuPRL12_Yb2Subradiant} is more elusive due to the inherently weak interaction with the environment.  Here we characterize the properties of deeply subradiant molecular states with intrinsic quality
factors exceeding $10^{13}$ via precise optical spectroscopy with the longest molecule-light coherent interaction times to date.  We find that two competing effects limit the lifetimes of
the subradiant molecules, with different asymptotic behaviors.  The first is radiative decay via weak magnetic-dipole and electric-quadrupole interactions.  We prove that its rate increases quadratically
with the bond length, confirming quantum mechanical predictions.  The
second is nonradiative decay through weak gyroscopic predissociation, with a rate proportional to the vibrational mode spacing and sensitive to short-range physics.  This work bridges the gap between atomic and molecular
metrology based on lattice-clock techniques \cite{KatoriNPhot11_LatticeClocks}, yielding new understanding
of long-range interatomic interactions and placing ultracold molecules at the forefront of precision measurements.
\end{abstract}
\date{\today}
\maketitle

Simple molecules provide a wealth of opportunities for precision measurements.  Their richer internal structure compared to atoms enables experiments that push the boundaries in determinations of the electric dipole moment of the electron \cite{ACMEScience14_ElectronEDM}, the electron-to-proton mass ratio and its variations \cite{SchillerBresselPRL12_HDIonMetrology,ChardonnetShelkovnikovPRL08_muStability}, and parity violation \cite{ChardonnetTokunagaMP13_ParityViolationMolecules}.  Diatomic molecules are moving to the forefront of many-body science \cite{YeYanNature13_KRbLatticeSpinModel} and quantum chemistry \cite{ZelevinskyMcGuyerPRL13_Sr2ZeemanNonadiabatic}, providing glimpses into fundamental laws \cite{UbachsDickensonPRL13_H2VibrPrecisMeast}.  However, this attractive complexity of molecular structure has historically posed difficulties for manipulation and modeling \cite{YeCarrNJP09_ColdMolecules}.  This work removes many of these barriers by employing techniques of optical lattice atomic clocks \cite{LudlowHinkleyScience13_10to18YbComparison,YeBloomNature14_10to18SrComparison} to control the quantum states of weakly bound homonuclear diatomic strontium molecules, in particular by using state-insensitive optical lattices \cite{YeSci08} for molecular transitions with three types of optical transition moments.  We observe strongly forbidden optical transitions in this asymptotic diatomic system, an ideal regime for studying the breakdown of the ubiquitous dipole approximation where the size of the quantum particle is a significant fraction of the resonant wavelength.  We explain these observations with a state-of-the-art ab initio molecular model \cite{MoszynskiSkomorowskiJCP12_Sr2Dynamics} and asymptotic scaling laws.
The results prove that the quantum mechanical effect of subradiance can be exploited for precision spectroscopy, and demonstrate the promise of combining precise state control, coherent manipulation, and accurate ab initio calculations with recently available ultracold molecular systems.

\begin{figure}
\includegraphics*[trim = 0in 0in 0in 0in, clip, width=3.5in]{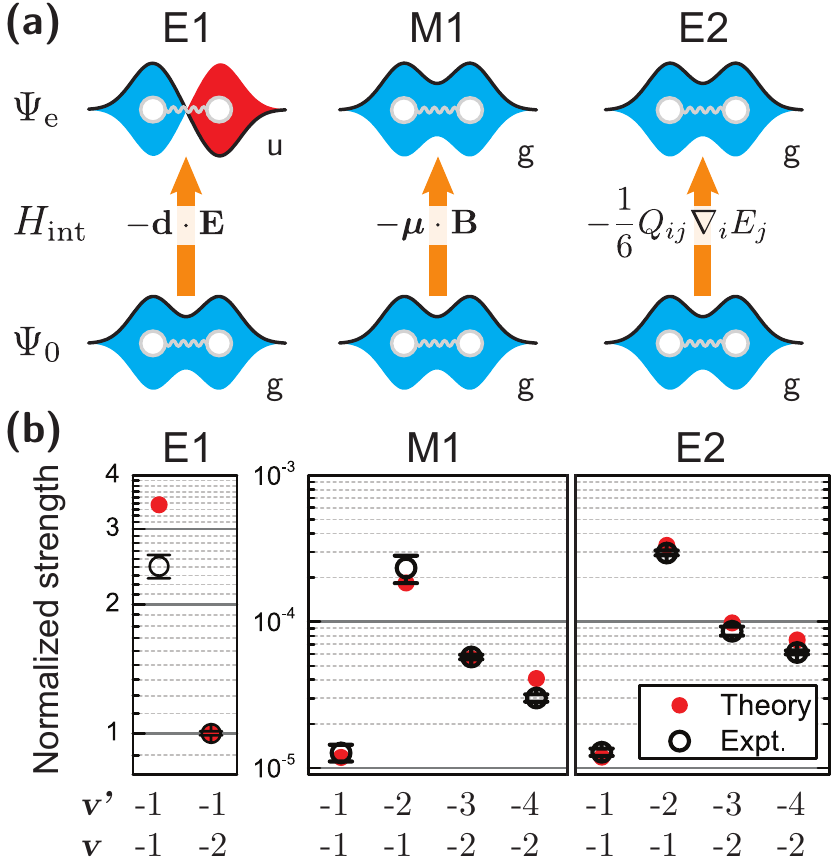}
\caption{\textbf{Optical transitions to superradiant and subradiant molecular states.}  \textbf{a,} Electric dipole (E1), magnetic dipole (M1), and electric quadrupole (E2) transitions in homonuclear diatomic molecules, from the gerade ground state to ungerade or gerade excited states.  \textbf{b,} Measurements and predictions of E1, M1, and E2 oscillator strengths in the weakly bound $^{88}$Sr$_2$ molecule. All values are normalized to the oscillator strength of an E1 transition to a superradiant $1_u$ level.  The error bars are standard errors of the mean of all $Q$ determinations.  For the M1 transition to $v'=-2$, the error bar was increased fourfold to reduce the difference between direct and Rabi-frequency measurements to two standard deviations.}
\label{fig:Fig1}
\end{figure}
We create Sr$_2$ molecules by photoassociation \cite{ZelevinskyReinaudiPRL12_Sr2} from an ultracold cloud of spinless strontium atoms, $^{88}$Sr, in an optical lattice satisfying the Lamb-Dicke and resolved-sideband conditions \cite{LeibfriedRMP03} (Methods).  The weak optical coupling of the ground $^1S_0$ state to the excited $^3P_1$ atomic state (22 $\mu$s lifetime \cite{ZelevinskyPRL06}) in Sr atoms enables spectroscopic resolution of molecular structure in the immediate proximity to the $^1S_0+{^3P_1}$ atomic threshold without losses from photon scattering.
This 689 nm intercombination (spin-changing) transition is electric-dipole (E1) allowed, where the photon couples states with opposite parity.  The magnetic dipole (M1) and electric quadrupole (E2) transitions are strictly forbidden.  Due to quantum mechanical symmetrization, these higher-order transitions become allowed in bound homonuclear dimers, as illustrated in Fig. 1a.  In the molecular ground state with the asymptotic electronic wavefunction $|\mathrm{X}^1\Sigma_g^+\rangle\approx|^1S_0\rangle|^1S_0\rangle$, only gerade (even) symmetry is possible, allowing optical E1 transitions only to ungerade (odd) excited molecular states.  However, M1 and E2 transitions are possible from X$^1\Sigma_g^+$ to gerade molecular states such as those near the $^1S_0+{^3P_1}$ threshold, since these higher moments couple states of the same symmetry.  Such transitions are very weak due to their spin- and electric-dipole-forbidden nature.  As a result, the gerade molecular states are subradiant, while the ungerade states are superradiant.  That is, if the E1 atomic radiative decay rate of $^3P_1$ to $^1S_0$ is $\Gamma$, then the equivalent rates are approximately $2\Gamma$ and $0$ for the superradiant and subradiant molecular states.  Asymptotically, these states correspond to the superpositions $\frac{1}{\sqrt{2}}(|^1S_0\rangle|^3P_1\rangle-|^3P_1\rangle|^1S_0\rangle)$ and $\frac{1}{\sqrt{2}}(|^1S_0\rangle|^3P_1\rangle+|^3P_1\rangle|^1S_0\rangle)$ of atomic states, respectively.  In this work, the subradiant states belong to the excited $1_g$ molecular potential (where ``1" refers to the total electronic angular momentum projection onto the molecular axis and ``$g$" to the symmetry of the electronic wavefunction), and couple to the ground state only via the higher-order M1 and E2 transitions.
We probe optical transition strengths to the subradiant molecular states to establish their asymptotic quadratic dependence on $R$, the classical expectation value of the bond length \cite{MoszynskiBusseryHonvaultMP06_Ca2AbInitio}.  This behavior is in stark contrast to the asymptotic E1 transition strengths of the superradiant states which are constant with $R$.

We have precisely quantified the optical transition oscillator strengths from X$^1\Sigma_g^+$ to the subradiant states.  The $1_g$ levels have vibrational quantum numbers $v'$ between $-1$ and $-4$ (counting from the continuum) and total angular momenta $J'=1,2$.  The oscillator strengths were measured via optical absorption spectra with areas normalized by the probe light power $P$ and pulse time $\tau$.  For each transition, the experimentally obtained quantity is $Q\equiv B_{12}/(c\pi^2w_0^2)=A/(\tau P)$, where $B_{12}$ is an Einstein $B$ coefficient, $w_0$ is the waist of the probe beam, and $A$ is the Lorentzian area of the natural logarithm of the absorption spectrum (Supplementary Information (SI)).  In Fig. 1b, the $Q$ values for the M1 and E2 transitions ($\Delta J=1$ and 2, respectively, all starting from a $J=0$ ground state) are normalized to the $Q$ for an E1 transition near the same atomic threshold, giving ratios of absorption oscillator strengths.  We find M1 and E2 $Q$ values that are 4-5 orders of magnitude suppressed compared to E1, as expected from the $Q\sim\frac{\pi^2}{4}\left(\frac{R}{\lambda}\right)^2$ ratio of the M1and E1 transition moments \cite{MoszynskiBusseryHonvaultMP06_Ca2AbInitio}.  Alternatively, oscillator strengths are proportional to the ratios of the squares of the Rabi frequencies to $P$, which were measured for M1 in the time domain by observing coherent Rabi oscillations (SI).  The two methods yield similar results.  We performed ab initio calculations of these doubly-forbidden transition strengths.  The results shown in Fig. 1b are in excellent agreement with measurements, confirming the asymptotic divergence of the M1 and E2 transition moments with $R$.  In the absence of this linear growth, the oscillator strengths would be governed by the rovibrational wavefunction overlaps (Franck-Condon factors), resulting in ratios different from our observations by about an order of magnitude.

\begin{figure}
\includegraphics*[trim = 0in 0in 0in 0in, clip, width=3.5in]{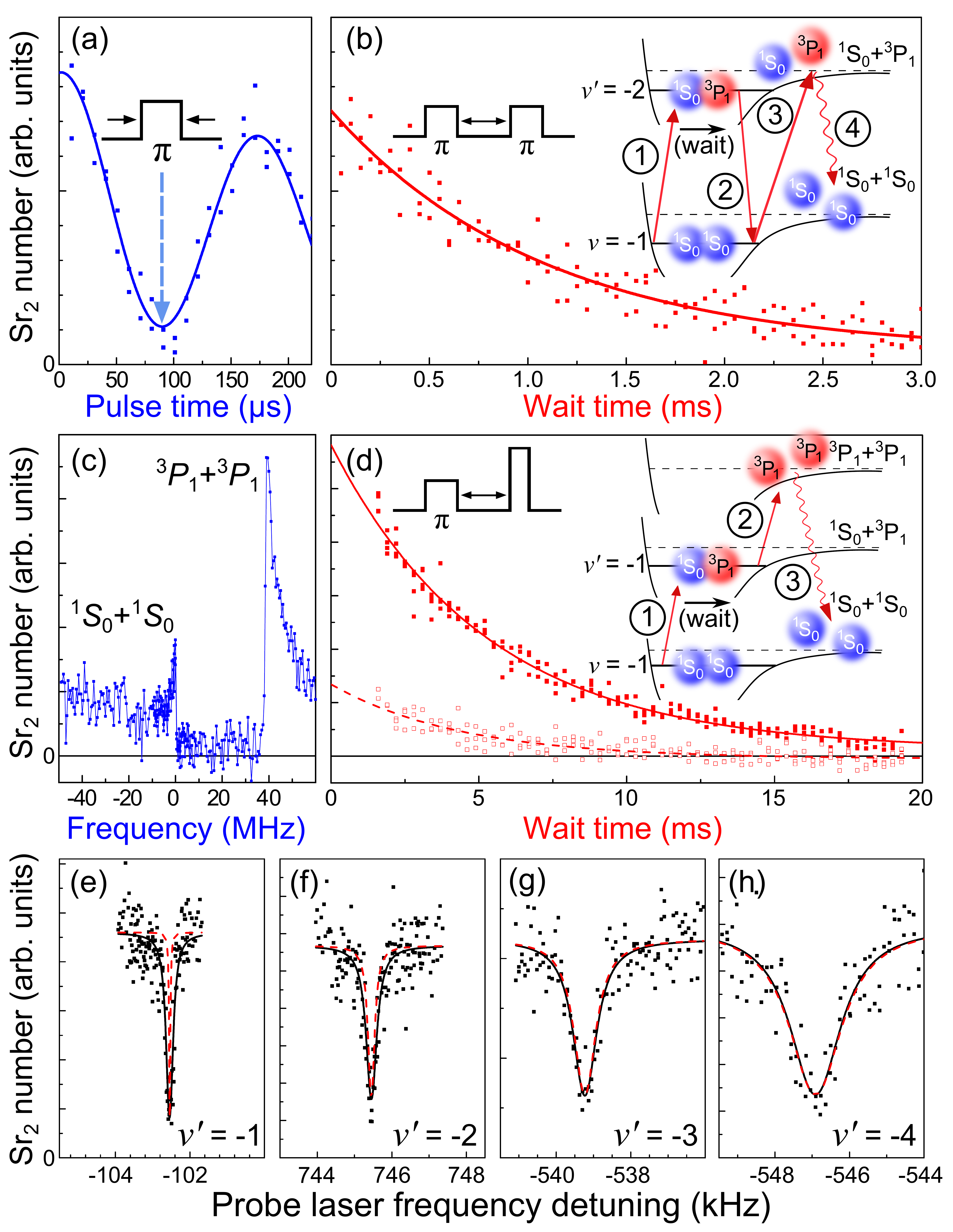}
\caption{\textbf{Direct and spectroscopic measurements of subradiant state lifetimes.}  For all states, $J'=1$.  \textbf{a,} Rabi oscillations between ground and excited gerade molecules that set the $\pi$-pulse lengths for lifetime measurements.  \textbf{b,} Excited-state population decay, fitted with an exponential curve.  The cartoon illustrates the four-step measurement sequence used for all the gerade states but the least-bound one.  \textbf{c,} The least-bound gerade state is strongly coupled to the atomic continuum due to its large bond length $R\approx130$ $a_0$.  The optical transition to the $^3P_1+{^3P_1}$ continuum corresponds to the right-hand peak (or shelf \cite{ZelevinskyMcGuyerPRL13_Sr2ZeemanNonadiabatic}), and the $^1S_0+{^1S_0}$ continuum to the left-hand oppositely facing peak.  \textbf{d,} Lifetime measurement of the least-bound state.  The cartoon shows the simplified measurement sequence, using the right peak in Fig. 2c.  The lower curve that was subtracted from the signal shows spontaneous fragmentation to the $^1S_0+{^1S_0}$ continuum during the wait time.  \textbf{e-h,} Optical spectra of the four $J'=1$ subradiant states, with their binding energies indicated in Table \ref{Tab:LineWidths}.  Dashed red lines indicate lineshapes deduced from direct lifetime measurements as in Fig. 2b,d.  Only the narrowest spectra are limited by technical broadening such as laser linewidth.}
\label{fig:Fig2}
\end{figure}
\begin{table}
\centering
\small
\begin{tabular}{ c | c r r c | r r r r }
   \multicolumn{5}{ c }{$\;\;\;\;\;\boldsymbol{J'=1}$} & \multicolumn{4}{ c }{$\boldsymbol{J'=3}$} \\
   \hline
   $v'$ & $E_b$ & $\gamma_{\rm{rad}}$ & $\gamma_{\rm{pre}}$ &$\gamma_{\rm{exp}}$ & $E_b$ & $\gamma_{\rm{rad}}$ & $\gamma_{\rm{pre}}$ & $\gamma_{\rm{exp}}$ \\
  \hline
 -1&19.0420(38)&5.7&19.7&28.5(2.0)&--&&& \\
 -2&316(1) &1.6&166&156.3(5.3)&193&1.7&819&$<$6E3 \\
 -3&1669(1) &0.8&555&525(30)&1438&0.9&3102&$<$13E3 \\
 -4&5168(1) &0.6&1243&1250(90)&4826&0.6&7033&$<$11E3 \\
\end{tabular}
\caption{\textbf{Measured and calculated contributions to the subradiant state linewidths.}  The binding energies $E_b$ are in MHz and the widths are in Hz.  The theory widths are ab initio.  The value and uncertainty for the $v'=-1$, $J'=1$ binding energy come from extrapolating a peak-to-shelf frequency difference \cite{ZelevinskyMcGuyerPRL13_Sr2ZeemanNonadiabatic} to zero magnetic field and probe and lattice light powers.}
\label{Tab:LineWidths}
\end{table}
We have also measured the lifetimes of the subradiant states.  The long molecule-light coherence times enable optical Rabi oscillations as shown in Fig. 2a, with the fringe decay times limited by the natural lifetimes of the $1_g$ states.  The Rabi period was used to determine the length of a $\pi$-pulse needed to excite the ground-state molecules into subradiant states.  After a variable wait time, the molecules were returned to the ground state and imaged via excitation to the $^1S_0+{^3P_1}$ continuum followed by spontaneous decay \cite{ZelevinskyReinaudiPRL12_Sr2}, as in the cartoon of Fig. 2b.  A typical exponential lifetime curve is shown in Fig. 2b.  While this approach was used for lifetime measurements of the $1_g$ states with $v'=-2,-3,-4$, the least-bound level allowed a simplified method.
The spectrum in Fig. 2c shows two bound-free optical transitions from $v'=-1$ to atomic continua. The process at the higher laser frequency corresponds to fragmentation via the doubly-excited $^3P_1+{^3P_1}$ continuum and is harnessed for direct lifetime measurements as depicted in Fig. 2d, where a plot of the recovered atom number versus wait time is shown with an exponential fit.
Even without an imaging pulse, some of the weakly-bound $v'=-1$ molecules decay to ground-state atoms, and we subtract this small contribution from the signal.
All known systematic effects were controlled (Methods).  The lifetime results are presented in Table \ref{Tab:LineWidths}.

Since the molecules are trapped in the Doppler-free regime, their absorption linewidths can also yield lifetimes.  Unlike direct lifetime measurements in Fig. 2b,d, this technique is sensitive to inhomogeneous broadening from stray magnetic fields and the lattice.  Therefore we engineered state-insensitive optical lattices for molecular transitions to the deeply subradiant states.  The polarization and wavelength were chosen to ensure light shifts $\lesssim1$ Hz$/$mW, leading to inhomogeneous broadening $<50$ Hz for 150 mW of lattice light power (SI and Fig. S1).
We nulled the ambient magnetic field to $\lesssim20$ mG by using the linear Zeeman effect in Sr$_2$, and applied a bias field of 0.43 G with angle control of $\lesssim2^{\circ}$ to define the quantization axis.
The four resulting spectra for transitions from X$^1\Sigma_g^+$ to $v'=-1,-2,-3,-4$ are shown in Fig. 2e-h, and are compared with lineshapes expected from direct lifetime measurements.  For the narrowest lines, the spectroscopic method overestimates the widths due to broadening caused by the intrinsic linewidth of the probe laser ($<200$ Hz), magnetic quenching ($<90$ Hz, Fig. 4), and the finite probe pulse ($<50$ Hz).

The radiative lifetimes of the $1_g$ states were calculated from the ab initio model by considering doubly-forbidden M1 and E2 transitions to the ground state.  The resulting contributions to the $\gamma_{\mathrm{rad}}$ linewidths are in the range of $\sim1$-$6$ Hz (Table \ref{Tab:LineWidths}).  Any contributions from decay to other states below the $^1S_0+{^3P_1}$ asymptote, as well as from black-body radiation \cite{FarleyPRA81_BBRRydbergLevels}, are negligible.  Unlike for atoms, the radiative lifetimes alone do not suffice to explain the observations.

\begin{figure}
\includegraphics*[trim = 0in 0in 0in 0in, clip, width=3.5in]{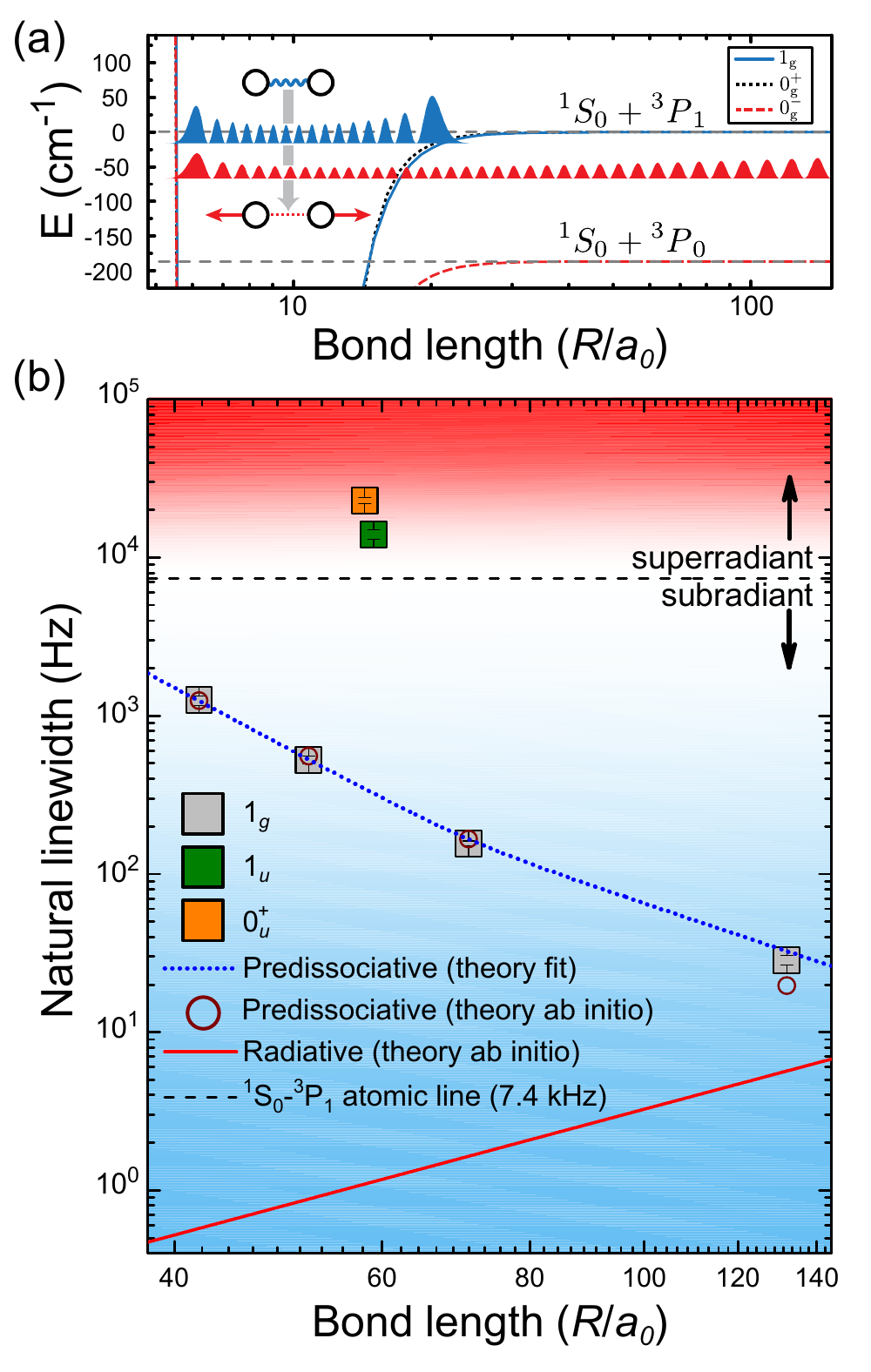}
\caption{\textbf{Natural linewidths of weakly bound subradiant and superradiant molecular states.} \textbf{a,} The dominant contribution to the natural lifetime of the long-range subradiant states is gyroscopic predissociation sensitive to short-range physics, as schematically shown here.  \textbf{b,}  Four least-bound subradiant states of $^{88}$Sr$_2$ with the lowest angular momentum $J'=1$ are measured, covering the range of bond lengths $R\sim40$-$130$ $a_0$.  The threshold between superradiant and subradiant behavior is marked, as well as measurements of two representative superradiant states.  Calculations of subradiant widths include both radiative and nonradiative contributions.  The former scale as $\propto R^2$ (the line is ab initio theory), and the latter as the vibrational energy spacing (the line is theory fit to the data with a single scaling parameter; ab initio theory points are also shown).  The error bars correspond to standard errors of the fitted exponential decay rates (as in Fig. 2b,d) for subradiant states or the Lorentzian widths (as in Fig. 2e-h) for superradiant states.  For $v'=-4$, the error bar is twice the standard error of the mean of the two measurement techniques to account for a larger statistical discrepancy.}
\label{fig:Fig3}
\end{figure}
Nonradiative decay is a dominant contributor to the subradiant lifetimes.  As shown in Fig. 3a, the $1_g$ bound states can couple to the long-lived $^1S_0+{^3P_0}$ continuum of the $0_g^-$ state.  The nature of this coupling is nonadiabatic Coriolis mixing \cite{JulienneMiesJCP84_MQDCouplingsDiatomics,MoszynskiSkomorowskiJCP12_Sr2Dynamics,ZelevinskyMcGuyerPRL13_Sr2ZeemanNonadiabatic} leading to weak gyroscopic predissociation.  An estimate of the predissociation rate follows from the Fermi golden rule, $2\pi\gamma_{\mathrm{pre}}\approx\frac{2\pi}{\hbar}|\langle1_g,v',J',m'|\hat{H}_R|0_g^-,E,J',m'\rangle|^2$, where $\hat{H}_R$ is the Coriolis interaction and $|0_g^-,E,J',m'\rangle$ are energy-normalized continuum scattering states with energy $E$.  This coupling vanishes at long range due to different dissociation thresholds of the $1_g$ and $0_g^-$ potentials, but not at short range (SI and Fig. S2).  We calculated the predissociative linewidths from the ab initio model, which was slightly tuned by scaling the $^3\Pi_g$ potential by $1.2\%$ to improve agreement with experiment.  Moreover, we can obtain accurate predissociative linewidth ratios without precise knowledge of the short-range physics.  The amplitude of a bound-state rovibrational wavefunction is $\psi_v(R)\propto\left(\frac{\partial E_v}{\partial v}\right)^{1/2}$, where $\frac{\partial E_v}{\partial v}$ is the known vibrational energy spacing \cite{MiesJCP84_MQDPredissociation} (SI).  Thus $\gamma_{\mathrm{pre}}=p\left(\frac{\partial E_v}{\partial v}\right)_{E=E_v}$, where the parameter $p$ can be related to the $^1S_0+{^3P_1}$ inelastic collision cross section \cite{JulienneMiesJCP84_MQDCouplingsDiatomics,IdoPRL05}.  The $\gamma_{\mathrm{pre}}$ values were obtained both from ab initio theory and by fitting $p=2.48\times10^{-7}$ to the measured $1_g$ level linewidths.

The results of the lifetime measurements and calculations are displayed in Fig. 3b, where the natural widths are shown versus $R$.  Note that our $R/\lambda\lesssim0.01$, which is less than $0.5\%$ of the range formerly explored with trapped ions \cite{DeVoePRL96_TwoIonsSubradSuperrad}.  The four $1_g$ subradiant states are marked, as well as two typical nearby superradiant states (from the $0_u^+$ and $1_u$ potentials).  The predictions for both nonradiative and radiative contributions are also shown.  The radiative contribution exhibits $\propto R^2$ asymptotic scaling.  The nonradiative contribution shows a change from roughly $\propto R^{-4}$ to $\propto R^{-2.5}$ scaling, reflecting the shift of long-range interaction from a $C_6$ to a $C_3$ character that occurs near $R\sim80$ $a_0$ for the $1_g$ potential of Sr$_2$ \cite{ZelevinskyPRL06}.  This scaling can be understood from the LeRoy-Bernstein formula \cite{BernsteinLeRoyJCP70_LongRangeDiatomicMoleculeScalings} relating the inverse density of states to the long-range $C_n/R^n$ behavior as $\frac{\partial E_v}{\partial v}\propto E_v^{(n+2)/(2n)}\propto R^{-(n+2)/2}$.

\begin{figure}
\includegraphics*[trim = 0in 0in 0in 0in, clip, width=3.5in]{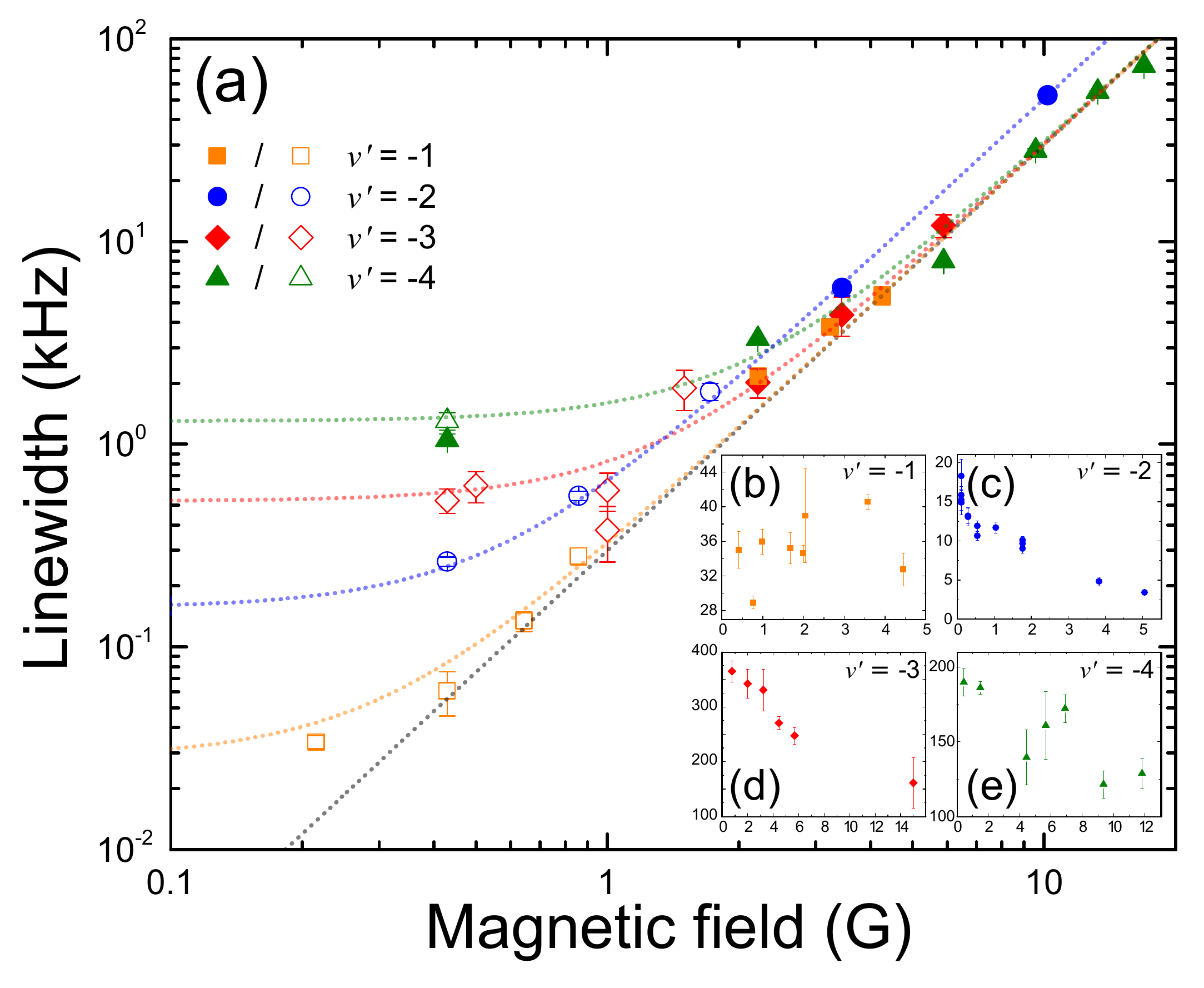}
\caption{\textbf{Magnetic-field tuning of subradiant lifetimes.}  The long-range subradiant states have linewidths that are highly tunable with small magnetic fields $B$.  \textbf{a,} The four linewidths of the transitions to $J'=1$ are shown versus $B$ in the range of $0.2$-$11$ G.  The low-field values are distinct (as in Fig. 3b), while at fields exceeding $\sim3$ G all widths increase quadratically with magnetic field at rates of $\sim300$-$500$ Hz$/$G$^2$.  The hollow points are from direct lifetime measurements, the filled points are spectroscopic linewidths, and the dotted lines are drawn to guide the eye and indicate zero-field widths.  \textbf{b-e,} The broader widths of the $J'=2$ partners decrease with applied field for $v'=-2,-3$.  All error bars are standard errors of the fitted Lorentzian widths or exponential decay rates.}
\label{fig:Fig4}
\end{figure}
Table \ref{Tab:LineWidths} summarizes the measurements and ab initio calculations for the $1_g$ levels.
Moreover, we found that the lifetimes of the subradiant states are tunable by orders of magnitude with modest magnetic fields up to $\sim10$ G.  Figure 4a shows the natural linewidths of the four $1_g$ states versus field strength.  They broaden with a quadratic coefficient of $\sim300$ Hz$/$G$^2$ (or $\sim500$ Hz$/$G$^2$ for $v'=-2$).  This broadening could be qualitatively explained via Zeeman mixing with nearby even-$J'$ levels that appear to be short-lived due to their more complex mixing dynamics, which is further substantiated by the narrowing trend of the $J'=2$ widths as shown in Fig. 4b-e.

The Sr$_2$ state with the narrowest natural linewidth ($v'=-1$) has a measured lifetime longer than that of the atomic $^3P_1$ state by an unprecedented factor of nearly 300, opening the door to ultrahigh-resolution molecular metrology.
Our precise determinations of the binding energies and Zeeman coefficients of molecular states in this deeply subradiant regime (SI and Table S1) should allow fine tuning of parameters in the ab initio molecular model to reach agreement with measurements at the experimental accuracy, which would be a major achievement of quantum chemistry.  Furthermore, Fig. 2c hints at the intriguing possibility of using long-lived states for ultracold molecule photodissociation \cite{LaneWellsPCCP11_CHLaserCooling}.  The shown transition from the least-bound subradiant excited state to the ground-state continuum should have an ultimate width limited by the subradiant state lifetime, corresponding to excess fragment energies of only a nanokelvin.

\section{Methods}

$^{88}$Sr atoms were laser-cooled in a two-stage magneto-optical trap (MOT) and loaded into a one-dimensional optical lattice with a depth of 30 $\mu$K and a wavelength near 900 nm.  The lattice was generated by a diode laser and semiconductor tapered amplifier, where a diffraction grating removed any amplified spontaneous-emission light.   Atoms were photoassociated into 3 $\mu$K molecules with a density of $\lesssim10^{12}/$cm$^3$, that were optically imaged by a photodissociation pulse with a high spectral resolution \cite{ZelevinskyReinaudiPRL12_Sr2}.
The molecules can be selectively created in either of the two least-bound vibrational levels ($v=-1$ or $-2$) of the electronic ground state.  They are distributed among two rotational levels with the total angular momentum $J=0$ or $2$, which are well resolved spectroscopically.  These molecules near the $^1S_0+{^1S_0}$ ground-state atomic threshold are the starting point for probing electronically excited molecules near the $^1S_0+{^3P_1}$ asymptote.
Narrow molecular transitions were induced with a laser that was phase-locked to the narrow-linewidth 689 nm cooling laser.  The trapping magnetic coils were pulsed off during spectroscopy, and other sources of magnetic field gradients and noise were eliminated.  For lifetime measurements, the following parameters were systematically controlled:  lattice light power, molecule density by adjusting the photoassociation light pulse detuning, magnetic field by adjusting current in a set of Helmholtz coils, and probe light power (for spectroscopic linewidth measurements).  No systematic shifts of the lifetime values were detected for the accessible densities and lattice intensities, that were each varied by roughly a factor of two.  Magnetic fields quench the lifetimes as in Fig. 4, so the ambient fields were carefully nulled.

The ab initio potentials for the $^3\Pi_g$ ($^1S+{^3P}$),  $^3\Sigma_g^+$
($^1S+{^3P}$), and $^1\Pi_g$ ($^1S+{^1P}$) electronic states (SI and Fig. S2) were calculated using linear response theory within the coupled-cluster singles and doubles framework.  The ground-state X$^1\Sigma_g^+$ empirical potential was used \cite{TiemannSteinEPJD10_Sr2XPotential}.  Excited-state potentials were fitted to analytical functions \cite{MoszynskiSkomorowskiJCP12_Sr2Dynamics}.  Spin-orbit couplings between the nonrelativistic states were fixed at their asymptotic values related to the atomic fine structure.  Rovibrational level calculations were set up in the Hund's case (a) framework by including the $^3\Pi_g$, $^3\Sigma_g^+$, and $^1\Pi_g$ electronic states for the $1_g$ symmetry, and $^3\Pi_g$ and $^3\Sigma_g^+$ states for the $0_g^-$ symmetry. Diagonalization of the multisurface Hamiltonian  for a given $J'$ was performed via the discrete variable representation method.

\section{Acknowledgments}

We gratefully acknowledge the NIST award 60NANB13D163 and the ARO grant W911NF-09-1-0504 for partial support of this work.  M. M. acknowledges the NSF IGERT DGE-1069260, and R. M. the Polish Ministry of Science and Higher Education for the grant NN204-215539.  R. M. also thanks the Foundation for Polish Science for support within the MISTRZ program.

\onecolumngrid
\appendix*

\section{Supplementary information}

\subsection{Engineering state-insensitive optical lattices for narrow molecular transitions}

\label{EngLLShift}
An optical lattice trap contributes no light shift or inhomogeneous broadening if the polarizabilties of the initial and final states for a transition are equal, $\alpha' = \alpha$.
In our measurements, we engineer such state-insensitive (``magic'') lattices for particular transitions by experimentally controlling both polarizabilities.

For states with total spin quantum number $J$ and azimuthal quantum number $m$ trapped by a linearly polarized lattice, the electric-dipole polarizability has the form \cite{AuzinshBudkerRochester}
\begin{align}
\alpha= \alpha_0(\lambda) + \alpha_2(\lambda) \,\left(\frac{3 \cos^2\theta-1}{2}\right) \left( \frac{3m^2-J(J+1)}{J(2J-1)} \right).
\label{eq:Alpha}
\end{align}
Equation (\ref{eq:Alpha}) emphasizes three experimentally accessible parameters:
(i) the wavelength $\lambda$ of the lattice light,
(ii) the angle $\theta$ of tilt between the directions of linear polarization for the lattice and of the quantization axis for the state, and
(iii) the choice of sublevel $m$.
For $\theta = 0$, it is equal to the standard ``$J$ representation'' in terms of the scalar and tensor polarizabilities $\alpha_0$ and $\alpha_2$, respectively \cite{ClarkMitroyJPB10_Polarizabilities}.
Note that $\alpha_2 = 0$ if $J<1$.

We experimentally control the angle $\theta$ for excited-state Sr$_2$ molecules by applying a magnetic field $B_z$ along the $z$-axis, perpendicular to the tight-trapping $x$-axis.
This field defines the quantization axis through the linear Zeeman interaction.
The angle $\theta$ is then set by a rotatable linear polarizer that controls the direction of the lattice polarization, which lies in the $yz$-plane.

\begin{figure}[h]
\includegraphics*[trim = 0in 0in 0in 0in, clip, width=6.5in]{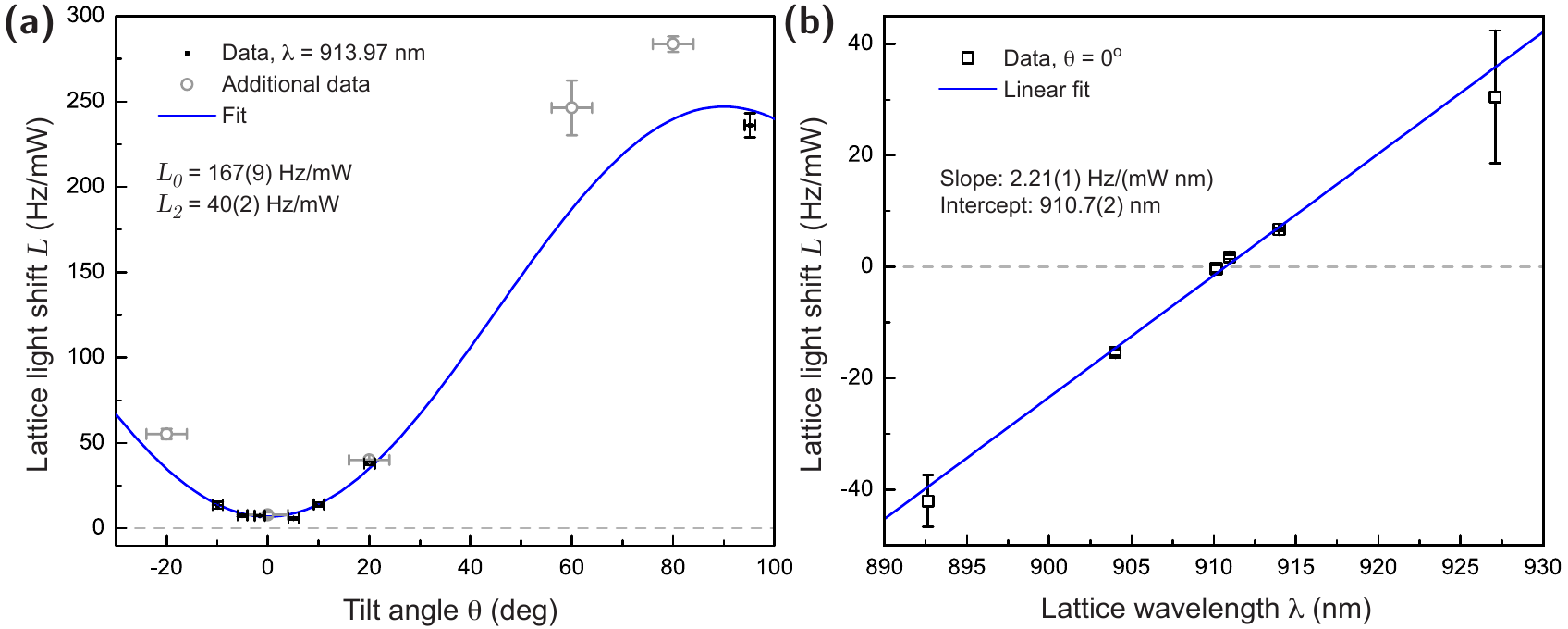}
\caption{\textbf{Polarization and wavelength tuning of the lattice light shift.}  The probed transition is from $v=-2$, $J=0$, $m=0$ of X$^1\Sigma_g^+$ to $v'=-4$, $J'=1$, $m'=0$ of $1_g$.  \textbf{a,} The shift is controlled by varying the lattice polarization direction relative to the quantum axis.  \textbf{b,} The shift is controlled by varying the lattice wavelength, at the optimal polarization determined from part (a).  The vertical error bars are standard errors from least-squares fitting of the slope $L$ of Eq. (\ref{Lcontrol}).  The horizontal error bars are instrumental uncertainties in setting the tilt angle.  The uncertainties in parenthesis for fit parameters shown are standard errors from least-squares fitting.}
\label{fig:AngleWavelength}
\end{figure}
For transitions from a $J=0$ ground state to an excited state with $J'$, we may use Eq. (\ref{eq:Alpha}) to introduce a lattice light-shift coefficient
\begin{align}	\label{Lcontrol}
L =\frac{1}{\hbar}\frac{\partial\Delta_{\mathrm{AC}}}{\partial P}=L_0(\lambda) + L_2(\lambda)  \left[ 3 \cos^2\theta - 1 \right] \left[ 3 (m')^2 - J'(J'+1) \right],
\end{align}
where $\Delta_{\mathrm{AC}}$ is the differential AC Stark shift, $P$ is the lattice light power, and the coefficients $L_0$ and $L_2$ depend on the choice of states.
This form highlights the experimental control of lattice light shifts.
To engineer a magic lattice for the transition, it is sufficient to select $m'$, $\theta$, and $\lambda$ such that $L = 0$.
A typical approach is sketched in Fig. \ref{fig:AngleWavelength}.
After choosing $m'$, the tilt angle $\theta$ is adjusted to minimize $L$, and the wavelength $\lambda$ is adjusted until $L=0$.
A similar approach was used to engineer a magic lattice for the $^1S_0$--$^3P_1$ E1 transition of $^{88}$Sr \cite{Ido3P1PRL03}, for which $\theta = \pi/2$ and $\lambda = 914(1)$ nm.
For M1 transitions to $1_g$ states with $J'=1$, we engineered magic lattices for $m'=0$ with $\theta = 0$ and $\lambda$ a few nm below 914 nm.
For E1 transitions to $0_u^+$ and $1_u$ states with $J'=1$, we engineered nearly magic lattices for $m'=0$ with $\theta = \pi/2$ but with $\lambda$ within 30 nm of the magic wavelength, because of laser limitations.  In this case, $|\alpha'/\alpha - 1| \lesssim 3\%$.  For the narrowest E2 transition to a $1_g$ state with $J'=2$, we engineered a nearly magic lattice for $m'=\pm2$ and $\theta=0$.

AC Stark shifts also contribute to inhomogeneous broadening of spectra in the optical lattice.  This broadening was estimated from measurements of the lattice light shift as $\partial\Gamma_{\mathrm{lat}}/\partial P\approx0.3|L|$ for our nearly magic lattice, where $\Gamma_{\mathrm{lat}}$ is the lattice contribution to the linewidth of a transition.  The factor of $\sim0.3$ comes from measurements of narrow transitions where the lattice uncertainty dominates the observed width \cite{ZelevinskyMcDonaldArxiv14_Sr2LatticeThermometry}.

\subsection{Transition strength measurements and calculations}

We measure a signal $S$ that is proportional to the number $n_J$ of ground-state Sr$_2$ ($v=-1$ or $-2$; $J=0$ or $2$) within a sampled volume of the optical lattice.
Prior to measurement, we apply a laser pulse with duration $\tau$ and power $P$ along the lattice axis.
Depending on the laser-frequency detuning from resonance, $\delta$, this laser pulse may induce absorption for a particular Sr$_2$ transition.
Our goal is to extract the strength of the transition probed by the laser pulse by analyzing $S(\delta)$, which is roughly a Lorentzian dip with a constant background.

We perform transition strength measurements from $J=0$ such that there is a unique sublevel $m=0$, but the procedure described below also applies to $J=2$.
During the laser pulse, the ground-state population $n_0(t)$ evolves as
\be
\frac{d}{dt}n_{0}(t)=-\Gamma_{0}(\delta)n_{0}(t),
\label{eq:DiffEq}
\ee
where $\Gamma_0(\delta)$ is the absorption rate per molecule.
After a pulse of duration $\tau$, the population is
\be
n_{0}(\tau)=n_{0}(0)\exp[-\Gamma_{0}(\delta)\tau ].
\ee
The measured signal is $S_0(\delta)\propto n_0(\tau)$, which far off-resonance we can denote with the shorthand $S_0(\infty)\propto n_0(0)$, such that
\be
\frac{S_0(\delta)}{S_0(\infty)}=\exp[-\Gamma_{0}(\delta)\tau].
\ee

We represent the transition strength by the quantity
\be
Q=-\frac{1}{\tau P}\int\ln\left[\frac{S_0(\delta)}{S_0(\infty)}\right]d\delta=\frac{1}{P}\int\Gamma_0(\delta)d\delta,
\label{Qdef}
\ee
because it is proportional to both an Einstein $B$ coefficient and an absorption oscillator strength as explained below.
No adjustments are made to account for the quantum numbers $m, J, m',$ or $J'$ in computing $Q$.
In practice, we find $Q$ by fitting a plot of $\ln S_0(\delta)$ with a Lorentzian of area $A$, which then gives $Q=A/(\tau P)$.
The default experimental units are MHz$/$(ms$\cdot\mu$W).
For both E1 and M1 transitions the area under one peak ($J=0, m = 0$ to $J'=1, m'=0$) was calculated.
For E2 transitions, the areas under two peaks ($J=0, m = 0$ to $J'=2, m'=\pm1$) were summed.
In all cases the probe laser light propagated along the tight-confinement $x$-axis of the lattice, perpendicular to the quantization $z$-axis set by an applied magnetic field.
For E1 and E2 transitions, the probe light was linearly polarized along $\hat{z}$, and for M1, along $\hat{y}$.  Area-preserving broadening mechanisms such as slightly state-sensitive optical lattice or magnetic-field variations do not affect $Q$, unlike area-changing mechanisms like power broadening.  Thus, care was taken to avoid power broadening.

The rate in Eq. (\ref{eq:DiffEq}) is \cite{HilbornAJP82_EinsteinCoeffsDipoleMomentsEtc}
\be
\Gamma_0(\delta)=\frac{W_{12}(\delta)}{N_1}\equiv\frac{1}{N_1}\int w_{12}(\omega)d\omega=B_{12}\int g(\omega)\rho(\omega)d\omega,
\label{eq:Gamma0}
\ee
where $W_{12}$ is an induced absorption rate, $N_1$ is the number of ground-state molecules, $\rho(\omega)$ is the probe laser energy density per angular frequency at $\delta$, $B_{12}$ is the Einstein $B$ coefficient of induced absorption, and $g(\omega)$ is a normalized lineshape function satisfying $\int g(\omega)d\omega=1$.
For narrow-linewidth lasers, Eq. (\ref{eq:Gamma0}) becomes
\be
\Gamma_{0}(\delta)=B_{12}g(2\pi\delta)I/c,
\ee
where for a probe beam waist $w_0$ and electric field amplitude $E$, the irradiance
\be
I=\frac{c\epsilon_0E^2}{2}=\frac{2P}{\pi w_0^2}.
\label{EtoP}
\ee
Then
\be
\int\Gamma_0(\delta)d\delta=\frac{B_{12}I}{2\pi c},
\label{eq:Gamma0B12}
\ee
and the quantity $Q$ we report is
\begin{align}
Q = \left( \frac{1}{ c \, \pi^2 w_0^2} \right) B_{12}
        = \left( \frac{1}{ c \, \pi^2 w_0^2} \right) \left( \frac{\pi e^2}{2 \epsilon_0 m_e \hbar \omega_{21}} \right) f_{12},
\label{eq:Q}
\end{align}
where $m_e$ and $e$ are the mass and charge of an electron and $\omega_{21}$ is the resonant angular frequency of the transition, which is nearly the same for all transitions considered.
As shown, this quantity is proportional to both $B_{12}$ and the dimensionless absorption oscillator strength $f_{12}$ for the transition under study.
The oscillator strength and the associated dipole and quadrupole operators are conventionally defined as in Ref. \cite{HasuoTojoPRA05_TotalReflecForbiddenTransEnhancement} and references therein.  For calculations, the operators are transformed into the molecular body-fixed frame.
Figure 1b in the manuscript is a plot of $Q$ values for the various studied transitions, each normalized by the $Q$ for one particular E1 transition.

Alternatively, $Q$ is related to Rabi frequency $\omega_R$ as follows.
Consider a two-level system $|1\rangle$ and $|2\rangle$ coupled by a time-dependent perturbation
\begin{align}
\hat{H}_\text{int}(t) = \hat{H}_0 \cos(\omega t).
\end{align}
For an E1 transition, $\hat{H}_0 = - \hat{{\bf d}}\cdot {\bf E}_0$, where $\hat{{\bf d}}$ is an electric dipole moment operator;
for M1, $\hat{H}_0 = - \hat{\boldsymbol{\mu}}\cdot {\bf B}_0$, where $\hat{\boldsymbol{\mu}}$ is a magnetic dipole moment operator;
for E2, $\hat{H}_0 = - (1/6) \hat{Q}_{ij} \nabla_i E_j$, were $\hat{Q}_{ij}$ is an electric quadrupole moment operator.
If the frequency $\omega=\omega_{21}$, then the population oscillates between the levels at the on-resonance Rabi frequency
\begin{align}
\omega_R =| \langle 1 |\hat{H}_0| 2 \rangle |/\hbar.
\end{align}
This Rabi frequency is related to the quantity $Q$ we report as
\begin{align}
Q = \frac{1}{4} \left(\frac{\omega_R^2}{P} \right)
\label{eq:QRabi}
\end{align}
and can be measured via Rabi oscillations.  Equation (\ref{eq:QRabi}) follows from Eqs. (\ref{EtoP}, \ref{eq:Gamma0B12}, \ref{eq:Q}) and $\int \Gamma_0(\delta) d\delta = (1/N_1) \int W_{12}(\delta) d\delta = (1/N_1)W_{12}(\omega_{21})\Gamma/4=\omega_R^2/4$, where $\Gamma$ is the transition linewidth, the last two steps follow Chs.~4 \& 5 of Ref. \cite{SiegmanLasersBook}, and we assumed a Lorentzian form of $W_{12}(\delta)$ and a narrow-linewidth laser.
Alternatively, Eq. (\ref{eq:QRabi}) may be derived for E1 transitions as in Ref. \cite{HilbornAJP82_EinsteinCoeffsDipoleMomentsEtc}.
Note that the factor of $1/4$ depends on the conventions used to define $w_0$ and $P$.
We confirmed this relationship for the M1 strengths that are plotted in Fig. 1b of the manuscript.

In calculations of oscillator strengths and radiative decay rates, we employed the following asymptotic forms for the non-vanishing components of the M1 and E2 transition moments between the X$^1\Sigma_g^+$ and $1_g$ electronic states \cite{MoszynskiBusseryHonvaultMP06_Ca2AbInitio},
\begin{align}
\label{eq:M1moment}
\langle \mathrm{X}^1\Sigma_g^+ |\hat{\mu}_{x/y}| 1_g \rangle &\approx \frac{\mu_B}{\sqrt{2}\hbar} R \langle ^1S_0 |\hat{p}_{x/y}|^3P_1\rangle, \\
\label{eq:E2moment}
\langle \mathrm{X}^1\Sigma_g^+ |\hat{Q}_{xz/yz}| 1_g \rangle &\approx \frac{1}{\sqrt{2}}
R \langle ^1S_0 |\hat{r}_{x/y}|^3P_1\rangle.
\end{align}
The growth of these transition moments with the bond length $R$ is explicit in Eqs. (\ref{eq:M1moment},\ref{eq:E2moment}).  Note that for E1 transitions to the $1_u$ and $0_u^+$ states, the long-range moment is constant with $R$,
\begin{align}
\langle \mathrm{X}^1\Sigma_g^+ |\hat{\bf{r}}| 1_u/0_u^+ \rangle \approx \sqrt{2} \langle ^1S_0 |\hat{\bf{r}}|^3P_1\rangle.
\label{eq:E1moment}
\end{align}

\subsection{Calculations of predissociative contributions to subradiant state lifetimes}

The main mechanism responsible for the finite lifetimes of the $1_g$ states is nonradiative decay (predissociation). The predissociation takes place via nonadiabatic coupling with states of the $0_g^-$ electronic potential
which correlates with the $^1S_0+{^3P_0}$ asymptote.  The coupling Hamiltonian
is of the form
\begin{equation}
\widehat{H}_R=-\frac{\hbar^2}{2\mu R^2}(\hat{J_+}{\hat
j_-}+\hat{J_-}{\hat j_+}),
\label{eq:HR}
\end{equation}
where $\hat{j}= \hat{L} + \hat{S} $, and $\hat{J}$,
$\hat{j}$, $\widehat{L}$ and $\hat{S}$ are the operators for the
total angular momentum, total electronic angular momentum, and electronic orbital and
spin angular momenta, respectively.

\begin{figure}[h]
\includegraphics*[trim = 0.5in 1in 0.5in 0.5in, clip, width=4.3in]{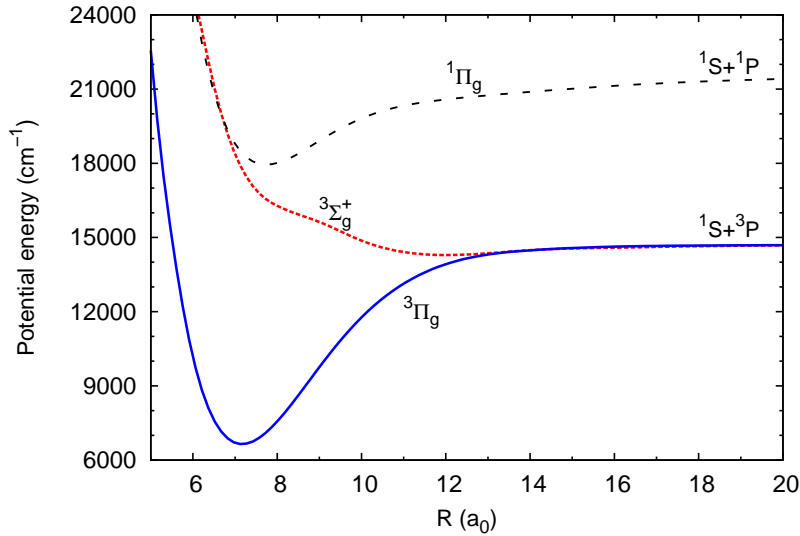}
\caption{\textbf{Potential energy curves for the $^3\Sigma_g^+$,  $^3\Pi_g$, and  $^1\Pi_g$  states
of Sr$_2$.}  This Hund's case (a) picture is helpful for analyzing short-range behavior.}
\label{fig:ShortRange}
\end{figure}
Predissociation widths were calculated from the Fermi golden rule,
\begin{equation}
  \Gamma_{pre}=\frac{2\pi}{\hbar}|\langle 1_g, v', J', m' |\widehat{H}_R|
  0_g^-, E, J', m' \rangle|^2,
  \label{eq:Fermi}
\end{equation}
where $| 0_g^-, E, J', m' \rangle$ is the energy-normalized continuum wavefunction
with energy $E$ matching that of the bound level $v'$.
We study subradiant states just below the atomic threshold, so it is reasonable to first analyze the asymptotic behavior of the coupling via $\widehat{H}_R$.  The nonadiabatic coupling in Eq.
(\ref{eq:Fermi}) vanishes at large interatomic separations because of the
different thresholds of the  $0_g^-$ and $1_g$ electronic
potentials, leading to
\begin{equation}
  \langle 1_g |\hat{j}_{\pm}| 0_g^-\rangle \approx  \langle ^3P_1 |\hat{j}_{\pm}|
  ^3P_0\rangle=0.
\end{equation}
However, the  nonadiabatic coupling does not vanish at short interatomic
separations.  To analyze the short-range behavior, it is convenient to use the Hund's
case (a) basis rather than (c).  In this picture, the $1_g$ state is mainly a mixture of $^3\Pi_g$ and  $^3\Sigma_g^+$ electronic states,
with a small contribution from $^1\Pi_g$.  The $0_g^-$ state is comprised
of the $^3\Pi_g$ and  $^3\Sigma_g^+$ states.  At short range,
$^3\Sigma_g^+$ is strongly repulsive and the $^3\Pi_g$ component
dominates, as shown in Fig. \ref{fig:ShortRange}.  Thus the nonadiabatic coupling from
Eqs. (\ref{eq:HR}, \ref{eq:Fermi}) will be  dominated by the following Coriolis interaction involving the $^3\Pi_g$ electronic components,
\begin{equation}
\langle 1_g, v', J', m' |\widehat{H}_R|0_g^-, E, J', m' \rangle\approx-
\langle  ^3\Pi_{g, |\Omega|=1}, v', J', m' |\frac{\hbar^2}{2\mu R^2}\left[\hat{J}_+{\hat
S}_-+\hat{J}_-{\hat S}_+\right]| ^3\Pi_{g, |\Omega|=0}, E, J', m'\rangle.
\label{Eq:HRRotationSpin}
\end{equation}
The explicit proportionality to $R^{-2}$ in Eq. (\ref{Eq:HRRotationSpin}) disappears after integration over the spatial coordinate, while the integral over the electronic spin and rotational degrees of freedom can be evaluated analytically,
\begin{equation}
\langle^3\Pi_{g,|\Omega|=1},J',m'|\hat{J}_+{\hat S}_-+\hat{J}_-{\hat S}_+|^3\Pi_{g,|\Omega|=0},J',m'\rangle=\sqrt{2J'(J'+1)}.
\end{equation}

We have calculated predissociation rates for weakly bound subradiant states
with $J'=1$ and $J'=3$, assuming Eq. (\ref{Eq:HRRotationSpin})
and that the Coriolis coupling is only relevant  at short interatomic distances
where the $^3\Pi_g$ and $^3\Sigma_g^+$ potentials are significantly different.  Due to the strong
oscillating character of rovibrational wavefunctions at short range, the
calculations of matrix elements in Eq. (\ref{eq:Fermi}) are difficult to converge.  The main
contribution to this integral comes from the range near the inner turning points
of the rovibrational wavefunctions.
To accurately reproduce the experimental linewidths without any
modification of the Sr$_2$ potentials, we require a better knowledge of the
short-range potentials than is currently available.
Therefore, to improve agreement of the theoretical and experimental linewidths for the $1_g$ $J'=1$ levels,
we scaled the $^3\Pi_g$ ab initio potential by $1.2\%$.

While it is challenging to obtain accurate ab initio values of the predisocciation
linewidths,
we can readily reproduce the width ratios for weakly bound levels by
using concepts from quantum defect theory to account for short-range effects.  A rovibrational wavefunction can be represented as \cite{MiesJCP84_MQDPredissociation}
\begin{equation}
  \psi_v(R)=\left(\frac{\partial E_v}{\partial v}\right)^{1/2}_{E=E_v}\left(\frac{2 \mu}{\pi \hbar^2}\right)^{1/2}
\alpha_v(R,k) \sin(\beta_v(R,k)),
\end{equation}
where $\frac{\partial E_v}{\partial v}$ is the vibrational spacing (alternatively,
$\frac{\partial v}{\partial E_v}$  is the density of states per unit energy), and
$\alpha_v(R,k)$ and $\beta_v(R,k)$ are the
quantum amplitude and phase of the state $v$. The $\alpha_v(R,k)$ and $\beta_v(R,k)$  functions depend on the local wavenumber $k(R)=\sqrt{2[E_v-V(R)]\mu}/\hbar$.
For different weakly bound levels, $\alpha_v(R,k)$ and $\beta_v(R,k)$ are nearly the same
at short range, so the $\Psi_{v'}$ wavefunctions differ only due to their local vibrational spacing.
The wavefunctions  $| 0_g^-, E, J', m'\rangle$  are nearly
identical for all $v'$ due to the large scattering energy $E$.
Thus, the matrix elements $\Gamma_{\mathrm{pre}}(v',J')$ from Eq. (\ref{eq:Fermi})
differ only due to the vibrational spacing factor.
Therefore, the predissociation rates can be written as
\begin{equation}
  \Gamma_{\mathrm{pre}}(v)= p\left(\frac{\partial E_v}{\partial v}\right)_{E=E_v},
  \label{eq:GammePreLinear}
\end{equation}
where the primes were dropped, and $p$ is the only free parameter quantifying the overlap between the scattering and bound rovibrational wavefunctions.

We can calculate $\frac{\partial E_v}{\partial v}$ by numerical
differentiation of the measured bound energies, and then
fit the single parameter $p$ to the measured widths as plotted in Fig. 3 of the manusript.
The linear dependence of the width on the vibrational spacing near the
asymptote, predicted in Ref. \cite{JulienneMiesJCP84_MQDCouplingsDiatomics}, is thus confirmed.
From the LeRoy-Bernstein formula \cite{BernsteinLeRoyJCP70_LongRangeDiatomicMoleculeScalings}
we can relate the vibrational spacing near the asymptote
to the bond length by assuming interatomic interaction of the form $C_n/R^n$,
\begin{equation}
  \frac{\partial E_v}{\partial v} \propto E_v^{\frac{n+2}{2n}} \propto
  R^{-\frac{n+2}{2}}.
\end{equation}
Thus the predissociation rate should scale $\propto1/R^{2.5}$ for the $C_3/R^3$ interaction and $\propto1/R^4$ for the $C_6/R^6$ interaction.  For the Sr$_2$ $1_g$ state, the $C_3$-$C_6$ crossover occurs near $v'=-2$, and our predissociation measurements are directly sensitive to it.

\subsection{Zeeman effect in subradiant states}

\begin{table}[h]
\centering
\small
\begin{tabular}{ccccccccc}
\hline
\hline
$J'$ & $v'$ & E$_{\mathrm{th}}$ & E$_{\mathrm{exp}}$ & $g_{\mathrm{th}}$ & $g_{\mathrm{exp}}$ & $|m'|$ & $\overline{q}_{\mathrm{th}}$ & $\overline{q}_{\mathrm{exp}}$ \\
\hline\hline
1	&	-1	&	19.3	&19.0420(38)	&	0.750	&0.749(1)		& 0 	&	-0.0865	&-0.086(2)	\\
 	&		&		&		&			&			& 1 	&	-0.0465	&-0.046(2)	\\
\hline
1	&	-2	&	315	&316(1)	&	0.750	&0.744(2)		& 0 	&	-0.0189	&-0.0192(3)\\
	&		&		&		&	 		&			& 1 	&	-0.0112	&-0.014(2)\\
\hline
1	&	-3	&	1651	&1669(1) &	0.750	&0.747(1)		& 0 	&	-0.00815	&-0.00817(6)\\
	&		&		&		&			&			& 1 	&	-0.00543	&-0.0062(4)\\
\hline
1	&	-4	&	5057	&5168(1) &	0.750	&0.747(1)		& 0 	&	-0.00543	&-0.00547(4)	\\
	&		&		&		&			&			& 1 	&	-0.00364	&-0.00472(5)		\\
\hline
2	&	-1	&	7.2	&7(1) &	0.429	&0.426(1)		& 0 	&	0.0288	&0.030(1)\\
	&		&		&		&			&			& 1 	&	0.0112	&0.014(1)\\
	&		&		&		&			&			& 2 	&	-0.0404	&-0.039(2)\\
\hline
2	&	-2	&	266	&270(1) &	0.347	&0.3480(4)	& 0 	&	0.00736	&0.0102(1)	\\
	&		&		&		&			&			& 1 	&	0.00364	&0.004(1)\\
	&		&		&		&			&			& 2 	&	-0.00729	&-0.009(1)\\
\hline
2	&	-3	&	1536	&1581(1) &	0.312	&0.308(2)		& 0 	&	0.00307	& \\
	&		&		&		&			&			& 1 	&	0.00157	&0.001(5)\\
	&		&		&		&			&			& 2 	&	-0.00300	&-0.001(2)\\
\hline
2	&	-4	&	4866	&5035(1) &	0.304	&0.293(1)		& 0 	&	0.0193	& \\
	&		&		&		&			&			& 1 	&	0.00093	&-0.001(1)\\
	&		&		&		&			&			& 2 	&	-0.00214	&-0.004(1)\\
\hline
3	&	-1	&	185	&193(1) & 0.125 &		&	& \\
\hline
3	&	-2	&	1401	&1438(1) & 0.125 &	&	& \\
\hline
3	&	-3	&	4684	&4826(1) & 0.125 &	&	& \\
\hline
\end{tabular}
\caption{\textbf{Predicted and measured Zeeman shifts for $1_g$ subradiant states of Sr$_2$.}  Coefficients up to the second order are included.  The binding energies $E$ are in MHz, the linear $g$-factors are unitless, and the quadratic shift coefficients $\overline{q}$ are in G$^{-1}$.  The $g$ and $q$ uncertainties are standard errors from least-squares fitting of spectroscopic peak positions versus magnetic field, and include the estimated inaccuracy of the applied magnetic field.  The value and uncertainty for the $v'=-1$, $J'=1$ binding energy come from extrapolating a peak-to-shelf frequency difference \cite{ZelevinskyMcGuyerPRL13_Sr2ZeemanNonadiabatic} to zero magnetic field and probe and lattice light powers.}
\label{Tab:Zeeman}
\end{table}
We have measured and calculated the linear and quadratic Zeeman-shift coefficients of the least-bound subradiant $1_g$ states with $J'=1,\;3$.  The results are shown in Table \ref{Tab:Zeeman}.
We parameterize linear and higher-order Zeeman shifts as \cite{ZelevinskyMcGuyerPRL13_Sr2ZeemanNonadiabatic}
\begin{align}	
\Delta E_b \approx g \mu_B m' B + \overline{q} \mu_B B^2
\end{align}
over a range of $\pm2$ G of the magnetic field $B$, where $\overline{q} = \overline{q}(v',J',m')$ depend on $m'$ while $g = g(v',J')$ do not.
Note that as defined, the binding energies are negative, so that positive shifts make molecules less bound.
For ideal Hund's case (c) $1_\text{g}$ states at the intercombination-line asymptote, the linear shifts \cite{ZelevinskyMcGuyerPRL13_Sr2ZeemanNonadiabatic} should have $g\approx(g_{\mathrm{atomic}})/ [J'(J'+1)]=3/4,\;1/8$ for $J' = 1,\;3$.

Comparison between the predicted and measured coefficients shows excellent agreement. Zeeman shift measurements are critical for refining the molecular model, since the linear coefficients constrain nonadiabatic Coriolis mixing between molecular states and the quadratic coefficients are highly sensitive to binding energies.
Table \ref{Tab:Zeeman} reveals nearly ideal Hund's case (c) shifts for $J'=1$, while for $J'=2$ the nonadiabatic contributions are large.  This may be explained by Coriolis mixing of the $1_g$ potential with $0_g^+$ (Fig. 3a in the manuscript).  Such mixing is not possible for odd $J'$, since in Sr$_2$ molecules the permutational symmetry permits only even-$J'$ levels for the $0_g^+$ potential.

\newpage

\begin{thebibliography}{37}
\expandafter\ifx\csname natexlab\endcsname\relax\def\natexlab#1{#1}\fi
\expandafter\ifx\csname bibnamefont\endcsname\relax
  \def\bibnamefont#1{#1}\fi
\expandafter\ifx\csname bibfnamefont\endcsname\relax
  \def\bibfnamefont#1{#1}\fi
\expandafter\ifx\csname citenamefont\endcsname\relax
  \def\citenamefont#1{#1}\fi
\expandafter\ifx\csname url\endcsname\relax
  \def\url#1{\texttt{#1}}\fi
\expandafter\ifx\csname urlprefix\endcsname\relax\def\urlprefix{URL }\fi
\providecommand{\bibinfo}[2]{#2}
\providecommand{\eprint}[2][]{\url{#2}}

\bibitem[{\citenamefont{Dicke}(1954)}]{DickePR54_CollectiveRadiation}
\bibinfo{author}{\bibfnamefont{R.~H.} \bibnamefont{Dicke}},
  \bibinfo{journal}{Phys. Rev.} \textbf{\bibinfo{volume}{93}},
  \bibinfo{pages}{99} (\bibinfo{year}{1954}).

\bibitem[{\citenamefont{Eberly}(1972)}]{EberlyAJP72_Superradiance}
\bibinfo{author}{\bibfnamefont{J.~H.} \bibnamefont{Eberly}},
  \bibinfo{journal}{Am. J. Phys.} \textbf{\bibinfo{volume}{40}},
  \bibinfo{pages}{1374} (\bibinfo{year}{1972}).

\bibitem[{\citenamefont{Gross and
  Haroche}(1982)}]{HarocheGrossPR82_Superradiance}
\bibinfo{author}{\bibfnamefont{M.}~\bibnamefont{Gross}} \bibnamefont{and}
  \bibinfo{author}{\bibfnamefont{S.}~\bibnamefont{Haroche}},
  \bibinfo{journal}{Phys. Rev.} \textbf{\bibinfo{volume}{93}},
  \bibinfo{pages}{301} (\bibinfo{year}{1982}).

\bibitem[{\citenamefont{DeVoe and
  Brewer}(1996)}]{DeVoePRL96_TwoIonsSubradSuperrad}
\bibinfo{author}{\bibfnamefont{R.~G.} \bibnamefont{DeVoe}} \bibnamefont{and}
  \bibinfo{author}{\bibfnamefont{R.~G.} \bibnamefont{Brewer}},
  \bibinfo{journal}{Phys. Rev. Lett.} \textbf{\bibinfo{volume}{76}},
  \bibinfo{pages}{2049} (\bibinfo{year}{1996}).

\bibitem[{\citenamefont{Zhou and
  Odom}(2011)}]{OdomZhouNNano11_SubradiantPlasmons}
\bibinfo{author}{\bibfnamefont{W.}~\bibnamefont{Zhou}} \bibnamefont{and}
  \bibinfo{author}{\bibfnamefont{T.~W.} \bibnamefont{Odom}},
  \bibinfo{journal}{Nature Nanotech.} \textbf{\bibinfo{volume}{6}},
  \bibinfo{pages}{423} (\bibinfo{year}{2011}).

\bibitem[{\citenamefont{Takasu et~al.}(2012)\citenamefont{Takasu, Saito,
  Takahashi, Borkowski, Ciury{\l}o, and
  Julienne}}]{TakahashiTakasuPRL12_Yb2Subradiant}
\bibinfo{author}{\bibfnamefont{Y.}~\bibnamefont{Takasu}},
  \bibinfo{author}{\bibfnamefont{Y.}~\bibnamefont{Saito}},
  \bibinfo{author}{\bibfnamefont{Y.}~\bibnamefont{Takahashi}},
  \bibinfo{author}{\bibfnamefont{M.}~\bibnamefont{Borkowski}},
  \bibinfo{author}{\bibfnamefont{R.}~\bibnamefont{Ciury{\l}o}},
  \bibnamefont{and} \bibinfo{author}{\bibfnamefont{P.~S.}
  \bibnamefont{Julienne}}, \bibinfo{journal}{Phys. Rev. Lett.}
  \textbf{\bibinfo{volume}{108}}, \bibinfo{pages}{173002}
  (\bibinfo{year}{2012}).

\bibitem[{\citenamefont{Katori}(2011)}]{KatoriNPhot11_LatticeClocks}
\bibinfo{author}{\bibfnamefont{H.}~\bibnamefont{Katori}},
  \bibinfo{journal}{Nature Photon.} \textbf{\bibinfo{volume}{5}},
  \bibinfo{pages}{203} (\bibinfo{year}{2011}).

\bibitem[{\citenamefont{{The ACME Collaboration: J. Baron}
  et~al.}(2014)\citenamefont{{The ACME Collaboration: J. Baron}, Campbell,
  DeMille, Doyle, Gabrielse, Gurevich, Hess, Hutzler, Kirilov, Kozyryev
  et~al.}}]{ACMEScience14_ElectronEDM}
\bibinfo{author}{\bibnamefont{{The ACME Collaboration: J. Baron}}},
  \bibinfo{author}{\bibfnamefont{W.~C.} \bibnamefont{Campbell}},
  \bibinfo{author}{\bibfnamefont{D.}~\bibnamefont{DeMille}},
  \bibinfo{author}{\bibfnamefont{J.~M.} \bibnamefont{Doyle}},
  \bibinfo{author}{\bibfnamefont{G.}~\bibnamefont{Gabrielse}},
  \bibinfo{author}{\bibfnamefont{Y.~V.} \bibnamefont{Gurevich}},
  \bibinfo{author}{\bibfnamefont{P.~W.} \bibnamefont{Hess}},
  \bibinfo{author}{\bibfnamefont{N.~R.} \bibnamefont{Hutzler}},
  \bibinfo{author}{\bibfnamefont{E.}~\bibnamefont{Kirilov}},
  \bibinfo{author}{\bibfnamefont{I.}~\bibnamefont{Kozyryev}},
  \bibnamefont{et~al.}, \bibinfo{journal}{Science}
  \textbf{\bibinfo{volume}{343}}, \bibinfo{pages}{269} (\bibinfo{year}{2014}).

\bibitem[{\citenamefont{Bressel et~al.}(2012)\citenamefont{Bressel, Borodin,
  Shen, Hansen, Ernsting, and Schiller}}]{SchillerBresselPRL12_HDIonMetrology}
\bibinfo{author}{\bibfnamefont{U.}~\bibnamefont{Bressel}},
  \bibinfo{author}{\bibfnamefont{A.}~\bibnamefont{Borodin}},
  \bibinfo{author}{\bibfnamefont{J.}~\bibnamefont{Shen}},
  \bibinfo{author}{\bibfnamefont{M.}~\bibnamefont{Hansen}},
  \bibinfo{author}{\bibfnamefont{I.}~\bibnamefont{Ernsting}}, \bibnamefont{and}
  \bibinfo{author}{\bibfnamefont{S.}~\bibnamefont{Schiller}},
  \bibinfo{journal}{Phys. Rev. Lett.} \textbf{\bibinfo{volume}{108}},
  \bibinfo{pages}{183003} (\bibinfo{year}{2012}).

\bibitem[{\citenamefont{Shelkovnikov et~al.}(2008)\citenamefont{Shelkovnikov,
  Butcher, Chardonnet, and
  Amy-Klein}}]{ChardonnetShelkovnikovPRL08_muStability}
\bibinfo{author}{\bibfnamefont{A.}~\bibnamefont{Shelkovnikov}},
  \bibinfo{author}{\bibfnamefont{R.~J.} \bibnamefont{Butcher}},
  \bibinfo{author}{\bibfnamefont{C.}~\bibnamefont{Chardonnet}},
  \bibnamefont{and}
  \bibinfo{author}{\bibfnamefont{A.}~\bibnamefont{Amy-Klein}},
  \bibinfo{journal}{Phys. Rev. Lett.} \textbf{\bibinfo{volume}{100}},
  \bibinfo{pages}{150801} (\bibinfo{year}{2008}).

\bibitem[{\citenamefont{Tokunaga et~al.}(2013)\citenamefont{Tokunaga,
  Stoeffler, Auguste, Shelkovnikov, Daussy, Amy-Klein, Chardonnet, and
  Darqui\'{e}}}]{ChardonnetTokunagaMP13_ParityViolationMolecules}
\bibinfo{author}{\bibfnamefont{S.~K.} \bibnamefont{Tokunaga}},
  \bibinfo{author}{\bibfnamefont{C.}~\bibnamefont{Stoeffler}},
  \bibinfo{author}{\bibfnamefont{F.}~\bibnamefont{Auguste}},
  \bibinfo{author}{\bibfnamefont{A.}~\bibnamefont{Shelkovnikov}},
  \bibinfo{author}{\bibfnamefont{C.}~\bibnamefont{Daussy}},
  \bibinfo{author}{\bibfnamefont{A.}~\bibnamefont{Amy-Klein}},
  \bibinfo{author}{\bibfnamefont{C.}~\bibnamefont{Chardonnet}},
  \bibnamefont{and}
  \bibinfo{author}{\bibfnamefont{B.}~\bibnamefont{Darqui\'{e}}},
  \bibinfo{journal}{Mol. Phys.} \textbf{\bibinfo{volume}{111}},
  \bibinfo{pages}{2363} (\bibinfo{year}{2013}).

\bibitem[{\citenamefont{Yan et~al.}(2013)\citenamefont{Yan, Moses, Gadway,
  Covey, Hazzard, Rey, Jin, and Ye}}]{YeYanNature13_KRbLatticeSpinModel}
\bibinfo{author}{\bibfnamefont{B.}~\bibnamefont{Yan}},
  \bibinfo{author}{\bibfnamefont{S.~A.} \bibnamefont{Moses}},
  \bibinfo{author}{\bibfnamefont{B.}~\bibnamefont{Gadway}},
  \bibinfo{author}{\bibfnamefont{J.~P.} \bibnamefont{Covey}},
  \bibinfo{author}{\bibfnamefont{K.~R.~A.} \bibnamefont{Hazzard}},
  \bibinfo{author}{\bibfnamefont{A.~M.} \bibnamefont{Rey}},
  \bibinfo{author}{\bibfnamefont{D.~S.} \bibnamefont{Jin}}, \bibnamefont{and}
  \bibinfo{author}{\bibfnamefont{J.}~\bibnamefont{Ye}},
  \bibinfo{journal}{Nature} \textbf{\bibinfo{volume}{501}},
  \bibinfo{pages}{521} (\bibinfo{year}{2013}).

\bibitem[{\citenamefont{McGuyer et~al.}(2013)\citenamefont{McGuyer, Osborn,
  McDonald, Reinaudi, Skomorowski, Moszynski, and
  Zelevinsky}}]{ZelevinskyMcGuyerPRL13_Sr2ZeemanNonadiabatic}
\bibinfo{author}{\bibfnamefont{B.~H.} \bibnamefont{McGuyer}},
  \bibinfo{author}{\bibfnamefont{C.~B.} \bibnamefont{Osborn}},
  \bibinfo{author}{\bibfnamefont{M.}~\bibnamefont{McDonald}},
  \bibinfo{author}{\bibfnamefont{G.}~\bibnamefont{Reinaudi}},
  \bibinfo{author}{\bibfnamefont{W.}~\bibnamefont{Skomorowski}},
  \bibinfo{author}{\bibfnamefont{R.}~\bibnamefont{Moszynski}},
  \bibnamefont{and}
  \bibinfo{author}{\bibfnamefont{T.}~\bibnamefont{Zelevinsky}},
  \bibinfo{journal}{Phys. Rev. Lett.} \textbf{\bibinfo{volume}{111}},
  \bibinfo{pages}{243003} (\bibinfo{year}{2013}).

\bibitem[{\citenamefont{Dickenson et~al.}(2013)\citenamefont{Dickenson, Niu,
  Salumbides, Komasa, Eikema, Pachucki, and
  Ubachs}}]{UbachsDickensonPRL13_H2VibrPrecisMeast}
\bibinfo{author}{\bibfnamefont{G.~D.} \bibnamefont{Dickenson}},
  \bibinfo{author}{\bibfnamefont{M.~L.} \bibnamefont{Niu}},
  \bibinfo{author}{\bibfnamefont{E.~J.} \bibnamefont{Salumbides}},
  \bibinfo{author}{\bibfnamefont{J.}~\bibnamefont{Komasa}},
  \bibinfo{author}{\bibfnamefont{K.~S.~E.} \bibnamefont{Eikema}},
  \bibinfo{author}{\bibfnamefont{K.}~\bibnamefont{Pachucki}}, \bibnamefont{and}
  \bibinfo{author}{\bibfnamefont{W.}~\bibnamefont{Ubachs}},
  \bibinfo{journal}{Phys. Rev. Lett.} \textbf{\bibinfo{volume}{110}},
  \bibinfo{pages}{193601} (\bibinfo{year}{2013}).

\bibitem[{\citenamefont{Carr et~al.}(2009)\citenamefont{Carr, DeMille, Krems,
  and Ye}}]{YeCarrNJP09_ColdMolecules}
\bibinfo{author}{\bibfnamefont{L.~D.} \bibnamefont{Carr}},
  \bibinfo{author}{\bibfnamefont{D.}~\bibnamefont{DeMille}},
  \bibinfo{author}{\bibfnamefont{R.~V.} \bibnamefont{Krems}}, \bibnamefont{and}
  \bibinfo{author}{\bibfnamefont{J.}~\bibnamefont{Ye}}, \bibinfo{journal}{New
  J. Phys.} \textbf{\bibinfo{volume}{11}}, \bibinfo{pages}{055049}
  (\bibinfo{year}{2009}).

\bibitem[{\citenamefont{Hinkley et~al.}(2013)\citenamefont{Hinkley, Sherman,
  Phillips, Schioppo, Lemke, Beloy, Pizzocaro, Oates, and
  Ludlow}}]{LudlowHinkleyScience13_10to18YbComparison}
\bibinfo{author}{\bibfnamefont{N.}~\bibnamefont{Hinkley}},
  \bibinfo{author}{\bibfnamefont{J.~A.} \bibnamefont{Sherman}},
  \bibinfo{author}{\bibfnamefont{N.~B.} \bibnamefont{Phillips}},
  \bibinfo{author}{\bibfnamefont{M.}~\bibnamefont{Schioppo}},
  \bibinfo{author}{\bibfnamefont{N.~D.} \bibnamefont{Lemke}},
  \bibinfo{author}{\bibfnamefont{K.}~\bibnamefont{Beloy}},
  \bibinfo{author}{\bibfnamefont{M.}~\bibnamefont{Pizzocaro}},
  \bibinfo{author}{\bibfnamefont{C.~W.} \bibnamefont{Oates}}, \bibnamefont{and}
  \bibinfo{author}{\bibfnamefont{A.~D.} \bibnamefont{Ludlow}},
  \bibinfo{journal}{Science} \textbf{\bibinfo{volume}{341}},
  \bibinfo{pages}{1215} (\bibinfo{year}{2013}).

\bibitem[{\citenamefont{Bloom et~al.}(2014)\citenamefont{Bloom, Nicholson,
  Williams, Campbell, Bishof, Zhang, Zhang, Bromley, and
  Ye}}]{YeBloomNature14_10to18SrComparison}
\bibinfo{author}{\bibfnamefont{B.~J.} \bibnamefont{Bloom}},
  \bibinfo{author}{\bibfnamefont{T.~L.} \bibnamefont{Nicholson}},
  \bibinfo{author}{\bibfnamefont{J.~R.} \bibnamefont{Williams}},
  \bibinfo{author}{\bibfnamefont{S.~L.} \bibnamefont{Campbell}},
  \bibinfo{author}{\bibfnamefont{M.}~\bibnamefont{Bishof}},
  \bibinfo{author}{\bibfnamefont{X.}~\bibnamefont{Zhang}},
  \bibinfo{author}{\bibfnamefont{W.}~\bibnamefont{Zhang}},
  \bibinfo{author}{\bibfnamefont{S.~L.} \bibnamefont{Bromley}},
  \bibnamefont{and} \bibinfo{author}{\bibfnamefont{J.}~\bibnamefont{Ye}},
  \bibinfo{journal}{Nature} \textbf{\bibinfo{volume}{506}}, \bibinfo{pages}{71}
  (\bibinfo{year}{2014}).

\bibitem[{\citenamefont{Ye et~al.}(2008)\citenamefont{Ye, Kimble, and
  Katori}}]{YeSci08}
\bibinfo{author}{\bibfnamefont{J.}~\bibnamefont{Ye}},
  \bibinfo{author}{\bibfnamefont{H.~J.} \bibnamefont{Kimble}},
  \bibnamefont{and} \bibinfo{author}{\bibfnamefont{H.}~\bibnamefont{Katori}},
  \bibinfo{journal}{Science} \textbf{\bibinfo{volume}{320}},
  \bibinfo{pages}{1734} (\bibinfo{year}{2008}).

\bibitem[{\citenamefont{Skomorowski et~al.}(2012)\citenamefont{Skomorowski,
  Paw{\l}owski, Koch, and Moszynski}}]{MoszynskiSkomorowskiJCP12_Sr2Dynamics}
\bibinfo{author}{\bibfnamefont{W.}~\bibnamefont{Skomorowski}},
  \bibinfo{author}{\bibfnamefont{F.}~\bibnamefont{Paw{\l}owski}},
  \bibinfo{author}{\bibfnamefont{C.~P.} \bibnamefont{Koch}}, \bibnamefont{and}
  \bibinfo{author}{\bibfnamefont{R.}~\bibnamefont{Moszynski}},
  \bibinfo{journal}{J. Chem. Phys.} \textbf{\bibinfo{volume}{136}},
  \bibinfo{pages}{194306} (\bibinfo{year}{2012}).

\bibitem[{\citenamefont{Reinaudi et~al.}(2012)\citenamefont{Reinaudi, Osborn,
  McDonald, Kotochigova, and Zelevinsky}}]{ZelevinskyReinaudiPRL12_Sr2}
\bibinfo{author}{\bibfnamefont{G.}~\bibnamefont{Reinaudi}},
  \bibinfo{author}{\bibfnamefont{C.~B.} \bibnamefont{Osborn}},
  \bibinfo{author}{\bibfnamefont{M.}~\bibnamefont{McDonald}},
  \bibinfo{author}{\bibfnamefont{S.}~\bibnamefont{Kotochigova}},
  \bibnamefont{and}
  \bibinfo{author}{\bibfnamefont{T.}~\bibnamefont{Zelevinsky}},
  \bibinfo{journal}{Phys. Rev. Lett.} \textbf{\bibinfo{volume}{109}},
  \bibinfo{pages}{115303} (\bibinfo{year}{2012}).

\bibitem[{\citenamefont{Leibfried et~al.}(2003)\citenamefont{Leibfried, Blatt,
  Monroe, and Wineland}}]{LeibfriedRMP03}
\bibinfo{author}{\bibfnamefont{D.}~\bibnamefont{Leibfried}},
  \bibinfo{author}{\bibfnamefont{R.}~\bibnamefont{Blatt}},
  \bibinfo{author}{\bibfnamefont{C.}~\bibnamefont{Monroe}}, \bibnamefont{and}
  \bibinfo{author}{\bibfnamefont{D.}~\bibnamefont{Wineland}},
  \bibinfo{journal}{Rev. Mod. Phys.} \textbf{\bibinfo{volume}{75}},
  \bibinfo{pages}{281} (\bibinfo{year}{2003}).

\bibitem[{\citenamefont{Zelevinsky et~al.}(2006)\citenamefont{Zelevinsky, Boyd,
  Ludlow, Ido, Ye, Ciury{\l}o, Naidon, and Julienne}}]{ZelevinskyPRL06}
\bibinfo{author}{\bibfnamefont{T.}~\bibnamefont{Zelevinsky}},
  \bibinfo{author}{\bibfnamefont{M.~M.} \bibnamefont{Boyd}},
  \bibinfo{author}{\bibfnamefont{A.~D.} \bibnamefont{Ludlow}},
  \bibinfo{author}{\bibfnamefont{T.}~\bibnamefont{Ido}},
  \bibinfo{author}{\bibfnamefont{J.}~\bibnamefont{Ye}},
  \bibinfo{author}{\bibfnamefont{R.}~\bibnamefont{Ciury{\l}o}},
  \bibinfo{author}{\bibfnamefont{P.}~\bibnamefont{Naidon}}, \bibnamefont{and}
  \bibinfo{author}{\bibfnamefont{P.~S.} \bibnamefont{Julienne}},
  \bibinfo{journal}{Phys. Rev. Lett.} \textbf{\bibinfo{volume}{96}},
  \bibinfo{pages}{203201} (\bibinfo{year}{2006}).

\bibitem[{\citenamefont{Bussery-Honvault and
  Moszynski}(2006)}]{MoszynskiBusseryHonvaultMP06_Ca2AbInitio}
\bibinfo{author}{\bibfnamefont{B.}~\bibnamefont{Bussery-Honvault}}
  \bibnamefont{and}
  \bibinfo{author}{\bibfnamefont{R.}~\bibnamefont{Moszynski}},
  \bibinfo{journal}{Mol. Phys.} \textbf{\bibinfo{volume}{104}},
  \bibinfo{pages}{2387} (\bibinfo{year}{2006}).

\bibitem[{\citenamefont{Farley and Wing}(1981)}]{FarleyPRA81_BBRRydbergLevels}
\bibinfo{author}{\bibfnamefont{J.~W.} \bibnamefont{Farley}} \bibnamefont{and}
  \bibinfo{author}{\bibfnamefont{W.~H.} \bibnamefont{Wing}},
  \bibinfo{journal}{Phys. Rev. A} \textbf{\bibinfo{volume}{23}},
  \bibinfo{pages}{2397} (\bibinfo{year}{1981}).

\bibitem[{\citenamefont{Mies and
  Julienne}(1984)}]{JulienneMiesJCP84_MQDCouplingsDiatomics}
\bibinfo{author}{\bibfnamefont{F.~H.} \bibnamefont{Mies}} \bibnamefont{and}
  \bibinfo{author}{\bibfnamefont{P.~S.} \bibnamefont{Julienne}},
  \bibinfo{journal}{J. Chem. Phys.} \textbf{\bibinfo{volume}{80}},
  \bibinfo{pages}{2526} (\bibinfo{year}{1984}).

\bibitem[{\citenamefont{Mies}(1984)}]{MiesJCP84_MQDPredissociation}
\bibinfo{author}{\bibfnamefont{F.~H.} \bibnamefont{Mies}}, \bibinfo{journal}{J.
  Chem. Phys.} \textbf{\bibinfo{volume}{80}}, \bibinfo{pages}{2514}
  (\bibinfo{year}{1984}).

\bibitem[{\citenamefont{Ido et~al.}(2005)\citenamefont{Ido, Loftus, Boyd,
  Ludlow, Holman, and Ye}}]{IdoPRL05}
\bibinfo{author}{\bibfnamefont{T.}~\bibnamefont{Ido}},
  \bibinfo{author}{\bibfnamefont{T.~H.} \bibnamefont{Loftus}},
  \bibinfo{author}{\bibfnamefont{M.~M.} \bibnamefont{Boyd}},
  \bibinfo{author}{\bibfnamefont{A.~D.} \bibnamefont{Ludlow}},
  \bibinfo{author}{\bibfnamefont{K.~W.} \bibnamefont{Holman}},
  \bibnamefont{and} \bibinfo{author}{\bibfnamefont{J.}~\bibnamefont{Ye}},
  \bibinfo{journal}{Phys. Rev. Lett.} \textbf{\bibinfo{volume}{94}},
  \bibinfo{pages}{153001} (\bibinfo{year}{2005}).

\bibitem[{\citenamefont{LeRoy and
  Bernstein}(1970)}]{BernsteinLeRoyJCP70_LongRangeDiatomicMoleculeScalings}
\bibinfo{author}{\bibfnamefont{R.~J.} \bibnamefont{LeRoy}} \bibnamefont{and}
  \bibinfo{author}{\bibfnamefont{R.~B.} \bibnamefont{Bernstein}},
  \bibinfo{journal}{J. Chem. Phys.} \textbf{\bibinfo{volume}{52}},
  \bibinfo{pages}{3869} (\bibinfo{year}{1970}).

\bibitem[{\citenamefont{Wells and Lane}(2011)}]{LaneWellsPCCP11_CHLaserCooling}
\bibinfo{author}{\bibfnamefont{N.}~\bibnamefont{Wells}} \bibnamefont{and}
  \bibinfo{author}{\bibfnamefont{I.~C.} \bibnamefont{Lane}},
  \bibinfo{journal}{Phys. Chem. Chem. Phys.} \textbf{\bibinfo{volume}{13}},
  \bibinfo{pages}{19036} (\bibinfo{year}{2011}).

\bibitem[{\citenamefont{Stein et~al.}(2010)\citenamefont{Stein, Kn\"{o}ckel,
  and Tiemann}}]{TiemannSteinEPJD10_Sr2XPotential}
\bibinfo{author}{\bibfnamefont{A.}~\bibnamefont{Stein}},
  \bibinfo{author}{\bibfnamefont{H.}~\bibnamefont{Kn\"{o}ckel}},
  \bibnamefont{and} \bibinfo{author}{\bibfnamefont{E.}~\bibnamefont{Tiemann}},
  \bibinfo{journal}{Eur. Phys. J. D} \textbf{\bibinfo{volume}{57}},
  \bibinfo{pages}{171} (\bibinfo{year}{2010}).

\bibitem[{\citenamefont{Auzinsh et~al.}(2010)\citenamefont{Auzinsh, Budker, and
  Rochester}}]{AuzinshBudkerRochester}
\bibinfo{author}{\bibfnamefont{M.}~\bibnamefont{Auzinsh}},
  \bibinfo{author}{\bibfnamefont{D.}~\bibnamefont{Budker}}, \bibnamefont{and}
  \bibinfo{author}{\bibfnamefont{S.}~\bibnamefont{Rochester}},
  \emph{\bibinfo{title}{{Optically Polarized Atoms: Understanding light-atom
  interactions}}} (\bibinfo{publisher}{Oxford University Press},
  \bibinfo{address}{Oxford}, \bibinfo{year}{2010}).

\bibitem[{\citenamefont{Mitroy et~al.}(2010)\citenamefont{Mitroy, Safronova,
  and Clark}}]{ClarkMitroyJPB10_Polarizabilities}
\bibinfo{author}{\bibfnamefont{J.}~\bibnamefont{Mitroy}},
  \bibinfo{author}{\bibfnamefont{M.~S.} \bibnamefont{Safronova}},
  \bibnamefont{and} \bibinfo{author}{\bibfnamefont{C.~W.} \bibnamefont{Clark}},
  \bibinfo{journal}{J. Phys. B} \textbf{\bibinfo{volume}{43}},
  \bibinfo{pages}{202001} (\bibinfo{year}{2010}).

\bibitem[{\citenamefont{Ido and Katori}(2003)}]{Ido3P1PRL03}
\bibinfo{author}{\bibfnamefont{T.}~\bibnamefont{Ido}} \bibnamefont{and}
  \bibinfo{author}{\bibfnamefont{H.}~\bibnamefont{Katori}},
  \bibinfo{journal}{Phys. Rev. Lett.} \textbf{\bibinfo{volume}{91}},
  \bibinfo{pages}{053001} (\bibinfo{year}{2003}).

\bibitem[{\citenamefont{McDonald et~al.}(2014)\citenamefont{McDonald, McGuyer,
  Iwata, and Zelevinsky}}]{ZelevinskyMcDonaldArxiv14_Sr2LatticeThermometry}
\bibinfo{author}{\bibfnamefont{M.}~\bibnamefont{McDonald}},
  \bibinfo{author}{\bibfnamefont{B.~H.} \bibnamefont{McGuyer}},
  \bibinfo{author}{\bibfnamefont{G.~Z.} \bibnamefont{Iwata}}, \bibnamefont{and}
  \bibinfo{author}{\bibfnamefont{T.}~\bibnamefont{Zelevinsky}},
  \bibinfo{journal}{arXiv:1409.5852}  (\bibinfo{year}{2014}).

\bibitem[{\citenamefont{Hilborn}(1982)}]{HilbornAJP82_EinsteinCoeffsDipoleMomentsEtc}
\bibinfo{author}{\bibfnamefont{R.~C.} \bibnamefont{Hilborn}},
  \bibinfo{journal}{Am. J. Phys.} \textbf{\bibinfo{volume}{50}},
  \bibinfo{pages}{982} (\bibinfo{year}{1982}).

\bibitem[{\citenamefont{Tojo and
  Hasuo}(2005)}]{HasuoTojoPRA05_TotalReflecForbiddenTransEnhancement}
\bibinfo{author}{\bibfnamefont{S.}~\bibnamefont{Tojo}} \bibnamefont{and}
  \bibinfo{author}{\bibfnamefont{M.}~\bibnamefont{Hasuo}},
  \bibinfo{journal}{Phys. Rev. A} \textbf{\bibinfo{volume}{71}},
  \bibinfo{pages}{012508} (\bibinfo{year}{2005}).

\bibitem[{\citenamefont{Siegman}(1986)}]{SiegmanLasersBook}
\bibinfo{author}{\bibfnamefont{A.~E.} \bibnamefont{Siegman}},
  \emph{\bibinfo{title}{{Lasers}}} (\bibinfo{publisher}{University Science
  Books}, \bibinfo{address}{California}, \bibinfo{year}{1986}).

\end{thebibliography}

\end{document}